\documentclass[sigconf]{acmart}
\acmConference[ESEC/FSE 2022]{The 30th ACM Joint European Software Engineering Conference and Symposium on the Foundations of Software Engineering}{14 - 18 November, 2022}{Singapore}

\usepackage{amsmath}
\usepackage{bm}
\usepackage{subcaption}
\usepackage{xcolor}
\usepackage{makecell,multirow} %
\usepackage{enumerate} 
\usepackage{setspace}
\usepackage{array}
\usepackage{booktabs}   
\usepackage{caption}
\usepackage{enumitem}
\usepackage{mathrsfs}
\usepackage{pifont}
\usepackage{tcolorbox}
\usepackage{cleveref}
\usepackage{graphicx}
\usepackage{tabularx}

\usepackage{mathcommand} %
\renewtextcommand\_{\textunderscore\allowbreak}%

\AtBeginDocument{
\crefformat{equation}{#2Equation (#1#3)}
\crefformat{section}{#2Section~#1#3}
\crefformat{subsection}{#2Section~#1#3}
\crefformat{subsubsection}{#2Section~#1#3}
\crefformat{algorithm}{#2Algorithm~#1#3}
\crefformat{figure}{#2Fig.~#1#3}
\crefformat{subfigure}{#2Fig.~#1#3}
\crefformat{table}{#2Table~#1#3}
}

\setcopyright{rightsretained}

\renewcommand\footnotetextcopyrightpermission[1]{}
\setcopyright{none}
\begin{document}

\title{Constructing Large-Scale Real-World Benchmark~Datasets~for~AIOps}

\settopmatter{authorsperrow=1}

\author{Zeyan Li$^{1}$, Nengwen Zhao$^{1}$, Shenglin Zhang$^{2}$, Yongqian Sun$^{2}$}
\author{Pengfei Chen$^{3}$, Xidao Wen$^{1}$, Minghua Ma$^{4}$, Dan Pei$^{1}$}

\affiliation{\institution{$^1$Tsinghua University,  $^2$Nankai University, $^3$Sun Yat-sen University, $^4$Microsoft Reasearch} \city{} \country{}}

\renewcommand{\shortauthors}{Z. Li, N. Zhao, S. Zhang, Y. Sun, P. Chen, X. Wen, M. Ma, D. Pei}

\begin{abstract}
   Recently, AIOps (Artificial Intelligence for IT Operations) has been well studied in academia and industry to enable automated and effective software service management. 
   Plenty of efforts have been dedicated to AIOps, including anomaly detection, root cause localization, incident management, etc. 
   However, most existing works are evaluated on private datasets, so their generality and real performance cannot be guaranteed. 
   The lack of public large-scale real-world datasets has prevented researchers and engineers from enhancing the development of AIOps. 
   To tackle this dilemma, in this work, we introduce three public real-world large-scale datasets about AIOps, mainly aiming at KPI anomaly detection, root cause localization on multi-dimensional data, and failure discovery and diagnosis.
   More importantly, we held three competitions in 2018/2019/2020 based on these datasets, attracting thousands of teams to participate.
   In the future, we will continue to publish more datasets and hold competitions to promote the development of AIOps further.

\end{abstract}

\maketitle

\section{Introduction}
In recent years, traditional manual operations and maintenance have gradually become time-consuming and error-prone due to the increasing complexity of modern IT service architectures. 
To tackle this dilemma, Artificial Intelligence for IT Operations (AIOps) has been proposed to relieve some manual intervention required. 
Gartner~\cite{gartner} defines AIOps as the application of machine learning (ML) and data mining to IT operations. 
AIOps combine big data and ML functionality to enhance and partially replace traditional manual efforts, including performance monitoring, anomaly detection, event correlation and analysis, incident management, etc.

In the literature, tremendous efforts have been devoted to the research of AIOps, including anomaly detection~\cite{bu2018rapid,donut,opprentice,ma2021jump}, root cause localization~\cite{li2019generic,gu2020efficient,ahmed2017detecting,bhagwan2014adtributor,jing2021autoroot,lin2016idice,lin2020fast,qureshi2020multidimensional,sun2018hotspot,wang2020impaptr,zhang2021halo,ma2020diagnosing}, failure prediction~\cite{ewarn,node,disk,outage,infocom_predict,tf_idf_lstm}, incident management ~\cite{fse_incident,assignment,chen2019continuous,lou2017experience}, resource management~\cite{delimitrou2014quasar,cortez2017resource} and so on.
However, due to the lack of public large-scale real-world datasets, it is difficult to compare these approaches fairly.
Although these works achieve promising performance on their respective private datasets, these approaches could perform unstably in the real world.
It is also difficult to transfer an approach to real generic application due to the various nature of different target systems for AIOps.
Thus, AIOps is at the crossing point between Artificial Intelligence research and operation engineers, which usually requires the cooperation of academic researchers and industry to obtain real data and scenarios and carry out research on this basis.
However, it is challenging for every researcher to find industrial partners since the monitoring data used for AIOps are highly confidential.
Therefore, the lack of public large-scale real-world datasets is one significant limitation of AIOps research and application.

Although there exist several public datasets, these datasets suffer from several limitations. 
First, the majority of existing datasets covers only one AIOps scenario, e.g., Yahoo Benchmark~\cite{egads} for KPI anomaly detection and \cite{zhou2019latent} for faulty microservice localization. 
However, AIOps involves diverse scenarios, for example, KPI anomaly detection, log anomaly detection, capacity planning, root cause localization, etc.
It is unsatisfactory to evaluate comprehensive AIOps approaches separately on such scenario-specific datasets.
Second, the scale of the dataset is limited. For example, Yahoo Benchmark for KPI anomaly detection only includes hundreds of data points. In general, real-world service systems tend to support numerous users and perform complex architecture. Thus, the dataset should be large and diverse to be close to the real world.  
Third, some datasets are synthetic, not from the real world. Such datasets cannot reflect the real system behavior, which could incur risk for real applications. 
As a result, rich, large-scale, and real-world datasets to assist the research of AIOps are highly desired.

To promote the development of AIOps, we have published three large-scale and real-world datasets, which are based on production systems of our industrial partner (e.g., Sougo, eBay, Tencent, Suning, and China Mobile Zhejiang).
These datasets are for KPI anomaly detection, root cause localization for multi-dimension data, and failure discovery and diagnosis, which is a comprehensive scenario, respectively.
There have been many research works~\cite{ren2019time,label-less,li2019generic,jing2021autoroot,cai2021modelcoder,yu2021microrank,li2021practical,chen2022adaptive} using our datasets.
Furthermore, based on these datasets, we have held an AIOps algorithm competition in 2018/2019/2020, attracting 125/141/141 teams and hundreds of participants\footnote{\url{https://competition.aiops-challenge.com/home/competition}}.
The solutions of top teams are also shared publicly~\cite{workshop}.
We believe such datasets would benefit the AIOps research and practice, as ImageNet~\cite{deng2009imagenet} did to the fields of computer vision and deep learning.
We would continue constructing and publishing new datasets and holding AIOps algorithm competitions once a year.

The main contributions of our work are summarized as follows.
\begin{itemize}[leftmargin=1em]
    \item Up to now, we have published three large-scale real-world datasets \cite{datasetA,datasetB,datasetC}, including KPI anomaly detection, root cause localization for multi-dimensional data, and failure discovery and diagnosis.
    These datasets have been used by many research works.
    \item We hold the yearly AIOps algorithm challenge, which involves thousands of researchers and engineers and significantly promotes this community's development.
\end{itemize}
\section{Our Work}
\label{sec:work}

Following existing work~\cite{notaro2021survey}, we categorize the research of AIOps into two macro-areas, failure management and resource management. 
Failure management aims to reduce the negative impact incurred by failures so as to ensure service availability and user experience. 
Resource management aims to allocate and save various resourcees (e.g., power, VMs) in IT service management. 
We currently mainly focus on failure management, and resource management could be our future work.

\begin{figure}[t]
    \captionsetup{skip=0pt}
    \centering
    \includegraphics[width=1\columnwidth]{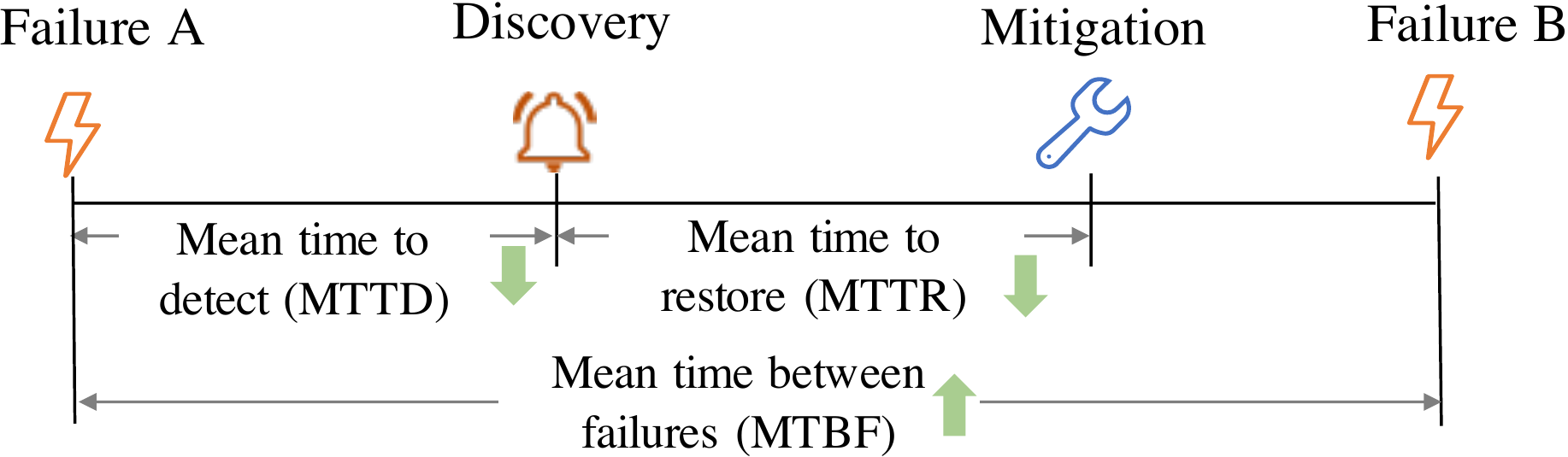}
    \caption{Pipeline and key measurements in failure management}
    \label{fig:time}
    \Description{The pipeline of failure management consists four stages}
\end{figure}

As presented in \cref{fig:time}, the goals of failure discovery, failure diagnosis,  and failure mitigation are to reduce the mean time to detect (MTTD), to reduce the mean time to recover (MTTR), and to increase the mean time between failures (MTBF), respectively. 
In our work, to tackle the bottleneck of AIOps research, motivated by ImageNet~\cite{deng2009imagenet}, we aim to construct large-scale real-world benchmark datasets for AIOps. 
We first focus on failure management and have published three related datasets, including KPI anomaly detection (\cref{sec:kpiad}), root-cause localization for multi-dimensional data (\cref{sec:mutli-dimensional-localization}), and failure discovery and diagnosis (\cref{sec:failure-discovery-and-diagnosis}). 
We will introduce each dataset and related background in detail in the following three sections. 

\section{KPI Anomaly Detection}
\label{sec:kpiad}
In this section, we first introduce the background of KPI anomaly detection. Then we present the details of our published dataset to enhance the research of KPI anomaly detection.

\subsection{Background} 
To closely monitor service running status and identify system failures, engineers tend to continuously collect many key performance indicators (KPIs) and identify anomalies from KPIs. 
Accurate KPI anomaly detection can trigger prompt troubleshooting, help to avoid economic loss, and maintain user experience~\cite{ren2019time,donut,opprentice,zhang2019cross}.

In recent years, tremendous efforts have been dedicated to KPI anomaly detection. 
Existing works could be classified into three categories, i.e., supervised methods (e.g., Opprentice~\cite{opprentice} and EDAGS~\cite{egads}), unsupervised methods (e.g., Donut~\cite{donut}), and semi-supervised methods (e.g., ADS~\cite{bu2018rapid}). 
However, there exist several significant challenges, restricting the research of KPI anomaly detection~\cite{ren2019time}. 
\begin{itemize}[leftmargin=1em]
    \item Various KPI patterns.  
    Based on our experience and observation, KPI patterns could be classified into three categories: seasonal (e.g., transaction volume), stable (e.g., success rate), and unstable (e.g., CPU utilization). 
     Therefore, a successful approach should be generalized enough for different patterns.

     \item Diverse KPI anomaly patterns. The definition of KPI anomalies is the behavior violating normal patterns, like spikes, dips, and jitters. Besides, different scenarios and components have different levels of tolerance to anomalies.
     Thus, the designed approaches should have the ability to detect diverse anomaly patterns. 
     
    \item Lack of sufficient labeled data. On the one hand, the number of KPIs is huge in large-scale systems. On the other hand, labeling anomalies requires experienced engineers with domain knowledge~\cite{label-less}. Thus, it is challenging and time-consuming to obtain sufficient labeled data to construct a supervised model and evaluate performance.  
    
\end{itemize}

Although there exist rich KPI anomaly detection works, these approaches are mainly evaluated on their private datasets. Besides, there are no large-scale real-world KPI anomaly detection datasets currently for model training and evaluation. Thus, despite the promising performance of their papers, the performance of real deployment in the industry is far from satisfying. 
Although there exist several public datasets about KPI anomaly detection like Yahoo Benchmark~\cite{egads} and Numenta Anomaly Benchmark~\cite{lavin2015evaluating}, they only contain KPIs with limited data points and synthetic anomalies. 
In comparison, the number and type of KPIs are large and diverse. 
Consequently, a large-scale and diverse KPI anomaly dataset is highly desired in the community of KPI anomaly detection.

\subsection{Dataset}
\label{sec:kpidataset}

\begin{figure}
\captionsetup{skip=0pt}
    \centering
    \includegraphics[width=1\columnwidth]{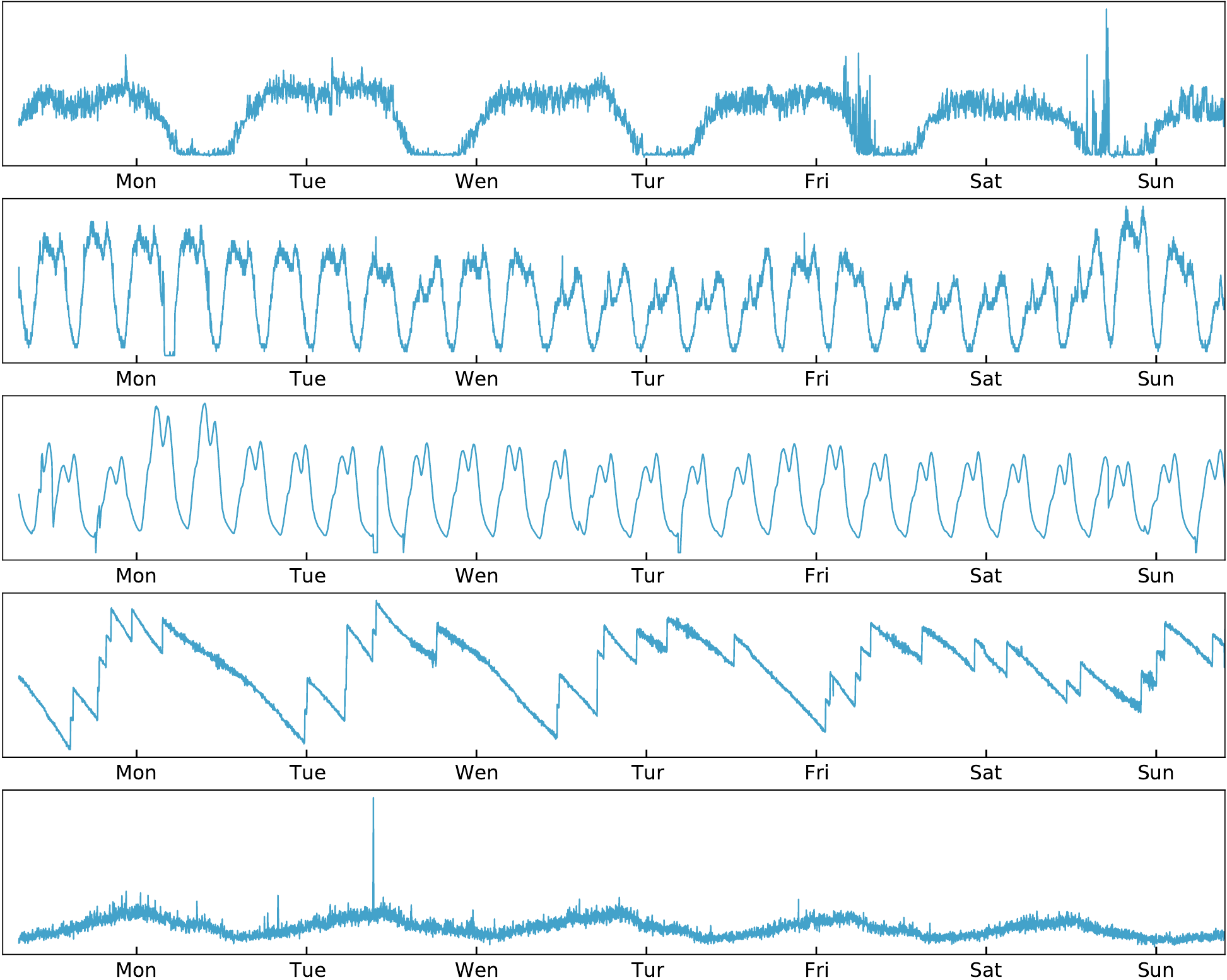}
    \caption{Some KPI examples of dataset $A$}
    \Description{Some KPI examples of dataset $A$}
    \label{fig:kpi}
    \Description{Some KPI examples of dataset $A$}
\end{figure}

To tackle the above challenges, we constructed a large-scale real-world KPI anomaly detection dataset (named $A$) with sufficient labels, covering various KPI patterns and anomaly patterns. 
This dataset is collected from five large Internet companies (Sougo, eBay, Baidu, Tencent, and Ali). Specifically, dataset $A$ includes 27 KPIs in total. For each KPI, experienced engineers in these companies manually labeled anomalies carefully. 
\Cref{tab:dataset} shows some statistical information about this dataset, including the total number of data points, anomaly ratio and time span. 
We could observe that the time span of each KPI is from two months to seven months, which is large-scale compared with existing datasets like Yahoo~\cite{egads}. 
\Cref{fig:kpi} presents five KPIs within a time span of one week. 
Obviously, the KPI patterns in this dataset are indeed various. 

\begin{table}
\captionsetup{skip=0pt}
\footnotesize
\caption{Overview of KPI anomaly detection dataset $A$}
\label{tab:dataset}
\centering
\begin{tabular}{cccc}
\toprule

    KPI ID    & \#Points & Anomaly ratio  & Time span (\#days) \\  \hline
    02e99bd4f6cfb33f & 241189 & 4.37\% & 183  \\ 
    046ec29ddf80d62e & 17568  & 0.45\% & 60   \\
    07927a9a18fa19ae & 21920  & 0.58\% & 86   \\ 
    09513ae3e75778a3 & 239975 & 0.09\% & 183  \\ 
    18fbb1d5a5dc099d & 240299 & 3.27\% & 183  \\ 
    1c35dbf57f55f5e4 & 240969 & 3.97\% & 183  \\ 
    40e25005ff8992bd & 209155 & 0.31\% & 146  \\ 
    54e8a140f6237526 & 16497  & 0.02\% & 61   \\ 
    71595dd7171f4540 & 295337 & 0.37\% & 207  \\ 
    769894baefea4e9e & 17568  & 0.05\% & 60   \\ 
    76f4550c43334374 & 17568  & 0.49\% & 60   \\ 
    7c189dd36f048a6c & 295379 & 0.14\% & 207  \\ 
    88cf3a776ba00e7c & 130872 & 2.37\% & 90   \\ 
    8a20c229e9860d0c & 17568  & 0.02\% & 60   \\ 
    8bef9af9a922e0b3 & 258907 & 0.20\% & 182  \\ 
    8c892e5525f3e491 & 294019 & 1.04\% & 207  \\ 
    9bd90500bfd11edb & 238798 & 0.05\% & 183  \\ 
    9ee5879409dccef9 & 130899 & 2.24\% & 90   \\ 
    a40b1df87e3f1c87 & 275850 & 0.13\% & 194  \\ 
    a5bf5d65261d859a & 237426 & 0.01\% & 183  \\ 
    affb01ca2b4f0b45 & 295361 & 0.19\% & 207  \\ 
    b3b2e6d1a791d63a & 16495  & 0.07\% & 61   \\ 
    c58bfcbacb2822d1 & 241453 & 0.05\% & 183  \\ 
    cff6d3c01e6a6bfa & 295258 & 0.36\% & 207  \\ 
    da403e4e3f87c9e0 & 241148 & 3.17\% & 183  \\ 
    e0770391decc44ce & 294048 & 1.04\% & 207  \\ \toprule
    \end{tabular}
    \end{table}

Based on this dataset, we held the first AIOps algorithm challenge in 2018\footnote{\url{https://competition.aiops-challenge.com/home/competition/1484452272200032281}}, attracting 125 teams and over 300 participants from both academia and industry. 
Besides, motivated by the novel evaluation strategy of KPI anomaly detection proposed in \cite{donut}, we import alarm delay restriction and design a novel evaluation method used in the competition\footnote{\url{https://github.com/iopsai/iops/tree/master/evaluation}}. This dataset and evaluation method have been widely used in many research works~\cite{ren2019time,label-less,chen2022adaptive}.
The best F1-score achieved in this competition is 0.8216. 
The solutions proposed by the top teams are shared in public~\cite{workshop}.

\section{Root Cause Localization for Multi-Dimensional Data}
\label{sec:mutli-dimensional-localization}

In this section, we first introduce the background of root cause localization for multi-dimensional data.
Then we introduce our published large-scale dataset for this problem.

\subsection{Background}
To closely monitor the events occurring in a system, various types of structured logs are generated, carrying critical information for failure diagnosis.
The structured logs record events with many fields describing the events.
For example, an online shopping platform could generate structured logs recording orders.
As shown in \cref{tbl:structured-logs-example}, the field \texttt{Timestamp} describes when the order occurs, \texttt{Dollar Amount} describes the dollar amount of the order, \texttt{Province} describes the location of the corresponding customer, and \texttt{ISP} describes the network that the customer used.
Engineers would like to collect the overall value of some fields, such as the total dollar amount, to monitor the status of the whole system.
We call such fields \textit{measures}~\cite{bhagwan2014adtributor,li2019generic}.
Other fields characterize the events and we call them \textit{attributes}~\cite{bhagwan2014adtributor,li2019generic,lin2016idice,gu2020efficient}.
By grouping original structured logs by some attributes and aggregating values of some fields, multi-dimensional data~\cite{li2019generic} are generated.
For example, to generate the multi-dimensional data in \cref{tbl:multi-dimensional-data-example}, we group structured logs like those in \cref{tbl:structured-logs-example} by \texttt{Timestamp}, \texttt{Province}, and \texttt{ISP}, and aggregate values of \texttt{Dollar Amount}.
Note that in \cref{tbl:multi-dimensional-data-example}, we only present a snapshot of the whole multi-dimensional data, i.e., the multi-dimensional data at a specific timestamp.
Multi-dimensional data is a more compact representation of the original structured logs, and thus, we focus on multi-dimensional data for failure diagnosis.

It indicates a failure occurring in the system when the overall value of a measure becomes abnormal (e.g., the total dollar amount decreases compared with the expected normal value).
However, in practice, a failure usually only causes the measure values under specific attribute combinations~\cite{gu2020efficient} abnormal~\cite{li2019generic}.
An \textit{attribute combination} refers to a slice of the whole multi-dimensional data by selecting records with specified values of some attributes, and it has been formally defined in existing literature~\cite{gu2020efficient}.
For example, in \cref{tbl:multi-dimensional-data-example}, only the measure values under \{\texttt{Province}=Beijing\} are significantly abnormal.
Such attribute combinations show the scope of the failure, and thus, are substantial clues for failure diagnosis.
In \cref{tbl:multi-dimensional-data-example}, for example, by localizing \{\texttt{Province}=Beijing\}, engineers get to know that the failure is related to the network or servers for users in Beijing, which directs further diagnosis.
Therefore, we call such attribute combinations as \textit{root causes of multi-dimensional data}.

\begin{table}[hbt]
\caption{Example structured logs}
\footnotesize
\begin{tabular}{ccccc}
\toprule
Order ID & Timestamp & Dollar Amount & Province & ISP \\ \midrule
T001 & 2021.11.11 10:00:01 & \$16 & Beijing & China Mobile \\ 
T002 & 2021.11.11 10:00:05 & \$21 & Beijing & China Unicom \\
\bottomrule
\end{tabular}
\label{tbl:structured-logs-example}
\end{table}

\begin{table}[hbt]
\footnotesize
\caption{Example multi-dimensional data}
\centering
\begin{tabular}{cccc}
\toprule
Province                                & ISP                                    & Dollar Amount                              & Expected Dollar Amount                            \\ \midrule
{\textbf{Beijing}} & {\textbf{China Mobile}} & {\textbf{5}} & {\textbf{10}} \\ 
{\textbf{Beijing}} & {\textbf{China Unicom}} & {\textbf{10}} & {\textbf{20}} \\ 
Shanghai                                & China Unicom                                 & 30                              & 31                               \\ 
Guangdong                               & China Mobile                                 & 10                                & 9.8                                 \\ 
Zhejiang                               & China Unicom                                 & 2                               & 2                                \\ 
Guangdong                               & China Unicom                                 & 200                               & 210                                \\ 
Shanxi                               & China Unicom                                 & 20                               & 22                                \\ 
Jiangsu                               & China Unicom                                 & 200                               & 203                                \\ 
Tianjin                               & China Mobile                                 & 41                               & 43                                \\ \hline
\multicolumn{2}{c}{Total} & 518 & 550.8\\ \bottomrule
\end{tabular}
\label{tbl:multi-dimensional-data-example}
\end{table}

\subsection{Dataset}
\label{sec:dataset-aiops-2019}
Most existing works~\cite{gu2020efficient,ahmed2017detecting,bhagwan2014adtributor,jing2021autoroot,lin2016idice,lin2020fast,qureshi2020multidimensional,sun2018hotspot,wang2020impaptr,zhang2021halo} on root cause localization for multi-dimensional data use their private datasets for evaluation.
Therefore, to enhance the research of root cause localization for multi-dimensional data, we construct and publish a dataset~\cite{datasetB}, named $B$, based on the real-world large online shopping platform of Suning. Ltd.

In dataset $B$, there are five attributes, which are named as \texttt{i}, \texttt{e}, \texttt{c}, \texttt{p} and \texttt{l} for confidential reasons.
The measure of $B$ is the average number of order per minute.
In total, the dataset $B$ contains 400 synthetic failures with ground-truth root cause labels, spanning eight weeks.
We use synthetic failures rather than real-world failures because it is hard to collect hundreds of production failures with ground-truth root causes.
The synthesis process of failures contains six steps.
\begin{enumerate}[leftmargin=1em]
    \item We first select a time point for anomaly injection where the overall measure values are smooth.
    \item We randomly choose $m\in\{1, 2, 3, 4\}$ different attributes out of the five attributes.
    \item Then we randomly choose $n\in\{1, 2, 3, 4\}$ different attribute combinations corresponding to the $m$ selected attributes. These $n$ selected attribute combinations are the ground-truth root cause where we are going to inject anomalies.
    \item For each selected attribute combination, we modify the measure values of it, i.e., the measure values of those rows at the select time point satisfying the attribute combination.
    The modified values are calculated by \textit{generalized ripple effect} (GRE)~\cite{li2019generic} with a random magnitude.
    \item We add extra noises, which are generated by Gaussian distribution, on both modified rows and other rows to simulate the cases where GRE does not work perfectly and there are forecast residuals, respectively.
    \item We check the synthetic failure with the following conditions. When anyone of them are triggered, we go back to step 2 to regenerate a failure at the time point.
    \begin{itemize}[leftmargin=1em]
        \item There is other attribute combinations sharing almost the same rows with the selected ground-truth attribute combinations.
        \item The injected noises is so large that the overall measure value of unmodified records (in step 3) is also abnormal.
    \end{itemize}
\end{enumerate}
Overall, among the 400 synthetic faults, 144/131/89/36 faults contains 1/2/3/4 root-cause attribute combinations, and 96/136/108/60 faults involve 1/2/3/4 attributes.

Based on the dataset $B$, we held the second AIOps challenge in 2019\footnote{\url{https://competition.aiops-challenge.com/home/competition/1484446614851493956}}.
There are 141 teams and 547 participants from both academia (39\%) and industries (61\%) in this competition.
The best performance achieved by them in this competition is 0.9593 with respect to F1-score.
The solutions proposed by the top teams are published in \cite{workshop}.

There are several existing research works~\cite{li2019generic,jing2021autoroot} utilizing the data provided by AIOps Challenge 2019.
Note that the datasets $\mathcal{B}_0\sim\mathcal{B}_4$ used and published by Squeeze~\cite{li2019generic} are based on exactly the same original data as the dataset $B$ described above and generated by almost the same synthesis process.

\section{Failure Discovery and Diagnosis}
\label{sec:failure-discovery-and-diagnosis}
In \cref{sec:kpiad} and \cref{sec:mutli-dimensional-localization}, we introduce our datasets on KPI anomaly detection and root cause localization for multi-dimensional data.
These datasets are useful to benchmark failure discovery and failure diagnosis algorithms respectively.
However, on the one hand, engineers require automatic failure discovery and diagnosis in an end-to-end manner.
On the other hand, root cause localization for multi-dimensional data provides only clues to the exact root causes, and it still requires further diagnosis to obtain fine-grained root causes given such clues.
Thus, we construct a dataset for benchmarking both failure discovery and fined-grained failure diagnosis approaches.
In this section, we first introduce the background of the problem.
Then we introduce the details of our dataset.

\subsection{Background}
\begin{figure}[tb]
\captionsetup{skip=0pt}
\centering
\includegraphics[width=0.9\columnwidth]{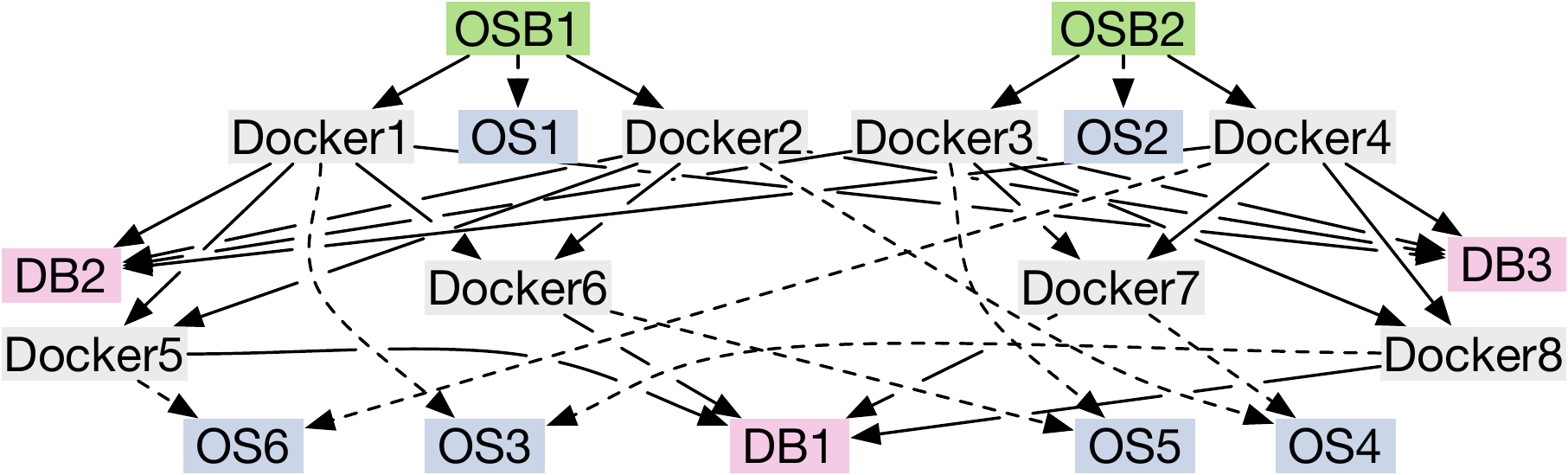}
\caption{The components and their relationships, including call dependencies (solid) and deployment dependencies (dashed), in the distributed system $\mathbb{S}$.
}
\label{fig:aiops-2020-sys-arch}
\Description{The components and their relationships, including call dependencies (solid) and deployment dependencies (dashed), in the distributed system $\mathbb{S}$.}   
\end{figure}

\begingroup
\setlength{\tabcolsep}{4pt} %
\begin{table*}[htb]
\captionsetup{skip=0pt}
\caption{A example trace in the dataset $C$}
\label{tbl:aiops-2020-trace-example}
\footnotesize
\begin{tabular}{ccccccccc}
\toprule
     callType & startTime & elapsedTime & success & traceId & id & pid & cmdbId & serviceName \\
     \midrule
    OSB & 1590249600016 & 274.0 & True & 6a171e24568385015da7 & 6a171919979385015da7 & None & os\_021 & osb\_001 \\
    CSF & 1590249600025 & 257.0 & True & 6a171e24568385015da7 & 6a1714eba31485915da7 & 6a171919979385015da7 & os\_021 & csf\_001\\
    RemoteProcess & 1590249600026 & 254.0 & True & 6a171e24568385015da7 & 6a17182dad5063a15da7 & 6a1714eba31485915da7 & docker\_003 & csf\_001\\
    CSF & 1590249600035 & 33.0 & True & 6a171e24568385015da7 & 6a171a703a9063325da7 & 6a17182dad5063a15da7&docker\_003&csf\_002\\
    RemoteProcess & 1590249600037&30.0&True& 6a171e24568385015da7 &6a1719334a3114525da7 & 6a171a703a9063325da7 & docker\_005&csf\_002\\
    LOCAL&1590249600054 & 7.0 & True & 6a171e24568385015da7 & 6a171648b90214635da7 & 6a1719334a3114525da7& docker\_005,& local\_method\_011 \\
    JDBC & 1590249600054 & 3.0 & True & 6a171e24568385015da7 & 6a171721a43214635da7 & 6a171648b90214635da7 & docker\_005 & db\_003\\
    \bottomrule
\end{tabular}    
\end{table*}
\endgroup

\begingroup\setlength{\tabcolsep}{4pt}
\begin{table*}[htb]
\captionsetup{skip=0pt}
\footnotesize
\caption{The metrics collected in the dataset $C$}
\label{tbl:aiops-2020-metrics}
\begin{tabularx}{2\columnwidth}{cX}
    \toprule
    \textbf{Category} & \textbf{Metrics list} \\ \midrule
    Docker & container\_\{thread\_total,fgct,thread\_idle,thread\_used\_pct,session\_used,thread\_running,fgc,cpu\_used,mem\_used, fail\_percent\}\\ \midrule
    Linux & Agent\_ping,Buffers\_used,CPU\_\{idle\_pct,user\_time,system\_time,iowait\_time,util\_pct\},Cache\_used,Disk\_\{rd\_ios,io\_util,await,avgqu\_sz,svctm,wr\_kbs,rd\_kbs,wr\_ios\},FS\_\{used\_space,used\_pct,total\_space,max\_util,max\_avail\},ICMP\_ping,Incoming\_network\_traffic,Memory\_\{used\_pct,used,available,free,available\_pct,total\},Num\_\{of\_processes,of\_running\_processes\},Outgoing\_network\_traffic,Page\_\{po,pi\},Processor\_\{load\_5\_min,load\_1\_min,load\_15\_min\},Received\_\{packets,errors\_packets,queue\},Recv\_total,Send\_total,Sent\_\{errors\_packets,queue,packets\},Shared\_memory,Swap\_used\_pct,System\_\{wait\_queue\_length,block\_queue\_length\},Zombie\_Process,ss\_total \\ \midrule
    Oracle & ACS,AIOS,AWS,Asm\_Free\_Tb,CPU\_\{free\_pct,Used\_Pct\},Call\_Per\_Sec,DFParaWrite\_Per\_Sec,DbFile\_Used\_Pct,DbTime,Exec\_Per\_Sec,Hang,LFParaWrite\_Per\_Sec,LFSync\_Per\_Sec,Logic\_Read\_Per\_Sec,Login\_Per\_Sec,MEM\_\{real\_util,Used,Total,Used\_Pct\},New\_\{Tbs\_Free\_Gb,Tbs\_Used\_Pct\},On\_Off\_State,PGA\_\{Used\_Pct,used\_total\},Physical\_Read\_Per\_Sec,Proc\_\{User\_Used\_Pct,Used\_Pct\},Redo\_Per\_Sec,Row\_Lock,SEQ\_Used\_Pct,SctRead\_Per\_Sec,SeqRead\_Per\_Sec,Sess\_\{Used\_Undo,Used\_Temp,Connect,Active\},Session\_pct,TPS\_Per\_Sec,Tbs\_\{Used\_Pct,Free\_Gb\},TempTbs\_Pct,Total\_Tbs\_Size,UndoTbs\_Pct,Used\_Tbs\_Size,User\_Commit,tnsping\_result\_time\\ \midrule
    Redis & Redis\_key\_count,blocked\_clients,connected\_clients,evicted\_keys,expired\_keys,instantaneous\_\{output\_kbps,input\_kbps,ops\_per\_sec\},keyspace\_\{hits,misses\},maxmemory,mem\_fragmentation\_ratio,redis\_\{ping,load\},rejected\_connections,total\_\{connections\_received,commands\_processed\},used\_\{memory\_peak,cpu\_sys,memory\_rss,cpu\_user,memory\} \\
    \bottomrule
\end{tabularx}    
\end{table*}
\endgroup

A distributed system is composed of many cooperating components, such as load balancers, web servers, application servers, and databases.
For example, in \cref{fig:aiops-2020-sys-arch}, we present the architecture of a production distributed system (named $\mathbb{S}$) at a major Internet service provider, China Mobile Zhejiang.
In \cref{fig:aiops-2020-sys-arch}, each vertex is a component.
These components are of five classes, including OSB (Oracle Service Bus), service, DB, Docker, and OS.
To monitor the status of the components, engineers collect various KPIs of these components, such as queries per minute, success rate, and average response time.

By KPI anomaly detection on these KPIs, we can discover failures in the distributed systems and figure out which components are abnormal.
However, due to the complex dependencies, a failure occurring at a root-cause component can propagate to other components and cause their KPIs abnormal as well~\cite{li2021practical}.
Therefore, to diagnose the failure, it is necessary to distinguish the root-cause components from other abnormal components.
Furthermore, to localize the fine-grained root causes of a failure, it is required to identify root-cause metrics of the root-cause components further.
For each component, there are plenty of metrics (e.g., CPU utilization and network throughput) collected.
When we identified the root cause metrics, for example, CPU usage, it is clear that the failure is caused by CPU exhaustion on the component.
In summary, we aim to discover failures by KPI anomaly detection and then identify the root-cause components and the root-cause metrics.

As introduced in \cref{sec:kpiad}, KPI anomaly detection has been extensively studied.
The existing approaches of identifying root-cause components~\cite{zhou2019latent, kim2013root, li2021practical, yu2021microrank, ma2020automap, lin2018microscope, gan2021sage} can be classified into two categories, i.e., trace-based~\cite{zhou2019latent,li2021practical,yu2021microrank,gan2021sage,lin2018microscope} and dependency graph-based~\cite{kim2013root,ma2020automap}.
There are also some works~\cite{liu2019fluxrank,wu2021microdiag} further identifying root-cause metrics.
MicroDiag~\cite{wu2021microdiag} models the dependencies among metrics by causality discovery and ranks metrics by PageRank.
FluxRank~\cite{liu2019fluxrank} ranks metrics by learning-to-rank.

\subsection{Dataset}

Most existing works on failure diagnosis use private datasets for evaluation.
Some existing works~\cite{zhou2019latent,li2021practical} published their datasets, but these datasets contain only traces and thus do not support fine-grained failure diagnosis.
Furthermore, though there are also some public datasets (e.g., Azure\footnote{https://github.com/Azure/AzurePublicDataset} and Alibaba\footnote{https://github.com/alibaba/clusterdata}) including fine-grained metrics, there are no ground-truth failure time points and root causes.
Therefore, to provide a comprehensive benchmark for failure discovery and fine-grained diagnosis, we construct and publish a dataset~\cite{datasetC}, named $C$.
Dataset $C$ has already been used by existing research works~\cite{cai2021modelcoder,yu2021microrank,li2021practical}.

The dataset $C$ is based on $sysA$, whose architecture is shown in \cref{fig:aiops-2020-sys-arch}.
There are three types of monitoring data in the dataset $C$, i.e., traces, KPIs,  and metrics. 
User requests are sent to one of the OSB components and then invoke other components, including Service and DB.
The whole execution process of a user request is called a trace.
We record each trace by its spans on each component, including the attributes of each span and the causal relationships among spans.
A span is a unit of work done in a component of a distributed system\footnote{We follow OpenTracing specification: \url{https://opentracing.io/docs/overview/spans/}}.
The attributes of the spans on each component can be aggregated into KPIs to reflect the overall status of each component.
In \cref{tbl:aiops-2020-trace-example}, we preset an example trace, where each row is a span.
Each span has a unique ID (the field \texttt{id}), the ID of its parent span (the span that invokes this span) (\texttt{pid}), and the unique ID of the whole trace (\texttt{traceId}).
With these three fields we can identify the spans of a trace and the causal relationships among the spans.
The other fields characterize a span.
For example, the span of the first row occurs at 1590249600016 (\texttt{startTime}, Unix timestamp in milliseconds), costs 274 milliseconds, and has a successful response.
Its call type is OSB and occurs at the component os\_021.
The KPIs (e.g., success rate, average response time) of the component are available by grouping the spans by \texttt{cmdbId} and aggregating some fields (e.g., \texttt{success}, \texttt{elapsedTime}).
Note that the overall KPIs (i.e., the KPIs on the OSB components) are directly given in the published dataset.
Besides the traces, we also collect the fine-grained metrics of each component.
There are four categories of metrics, which are summarized in \cref{tbl:aiops-2020-metrics}.

In the dataset $C$, we provide 169 injected failures with ground-truth time points and root causes, spanning one month.
We use injected failures rather than real-world failures because it is hard to collect so many real-world failures with ground-truth root causes.
On $sysA$, we injected 7 types of failures in total, which are summarized in \cref{tbl:aiops-2020-failure-injection}.

\begingroup
\setlength{\tabcolsep}{2pt} %
\begin{table}[htb]
\captionsetup{skip=0pt}
\caption{Failure injection in the dataset $C$}
\label{tbl:aiops-2020-failure-injection}
\footnotesize
\begin{tabular}{lll}
\toprule
    Component & Injection type & Description\\ \midrule
    Database & close & Close the listening port of the target instances\\
    Database & session limit & Decrease the session limit of the target instance\\
    Container & CPU stress~* & Stress the CPU of the target container\\
    Container & network delay~\$ & Delay packets randomly on the target container\\
    Container & network loss~\$ & Drop packets randomly on the target container\\
    Physical node & network delay~\$ & Delay packets randomly on the target node\\
    Physical node & network loss~\$ & Drop packets randomly on the target node\\
    \bottomrule
    \multicolumn{3}{l}{Injection tool: \texttt{stress-ng} (*) and \texttt{tc} (\$)}
\end{tabular}
\end{table}
\endgroup

Based on the dataset $C$, we held the third AIOps Challenge in 2020\footnote{\url{https://competition.aiops-challenge.com/home/competition/1484441527290765368}}.
It attracted 141 teams and 517 participants from both academia and industries.
In this competition, the participants are asked to detect and diagnose these failures in an online manner.
For each failure, each team is allowed to submit at most two potential root-cause metrics.
For network failures on containers, including both loss and delay, as we did not collect network metrics on containers, the participants are asked to localize the root-cause containers only.
If the result submitted by a team of a failure achieves a precision score greater than or equal to 0.5, then the result is considered valid.
For each failure, the valid results from different teams are ranked by $\text{time to diagnose} / \text{F-0.5 score}$, then each of them gets $\max(10 - i + 1, 0)$ points ($i$ is the rank). 
The best score achieved by the participants is 755, given 129 failures in total.
\section{Conclusion}

AIOps has attracted a great deal of attention from academics and industry. 
A significant limitation of the research of AIOps is the lack of public real-world and large-scale datasets. 
To tackle this problem, we have published three datasets, including KPI anomaly detection, root-cause localization for multi-dimensional data, failure discovery, and diagnosis. 
More importantly, we held an algorithm competition once a year based on the public datasets, attracting hundreds of teams to take part. 
Our work is helpful for practitioners and researchers to apply AIOps to enhance service reliability.
In the future, we will continuously publish more datasets involving various AIOps scenarios and hold the competition. 
Welcome to pay attention to and actively participate in the competition.

\clearpage{}
\bibliographystyle{ACM-Reference-Format}
\balance
\bibliography{refer} 
\end{document}